# Electrically driven exciton-polaritons in metal halide perovskite metatransistors

*Yutao Wang,[1,2][†] Jingyi Tian,[1,3][†] Maciej Klein,[1,3] Giorgio Adamo,[1,3] Son Tung Ha,[4][*] and Cesare Soci[1,3][*]*

[1] Centre for Disruptive Photonic Technologies, TPI, Nanyang Technological University, 21 Nanyang Link, Singapore 637371

[2] Interdisciplinary Graduate School, Energy Research Institute @NTU (ERI@N), Nanyang Technological University, 50 Nanyang Drive, Singapore 637553

[3] Division of Physics and Applied Physics, School of Physical and Mathematical Sciences, Nanyang Technological University, 21 Nanyang Link, Singapore 637371

[4] Institute of Materials Research and Engineering, Agency for Science Technology and Research (A*STAR), 2 Fusionopolis Way, Singapore 138634

[†]These authors contributed equally to this work

*Correspondence: Ha_Son_Tung@imre.a-star.edu.sg; csoci@ntu.edu.sg*

**Achieving electrical injection of exciton-polaritons, half-light, half-matter quasiparticles arising from the strong coupling between photonic and excitonic resonances, is a crucial milestone to scale up polaritonic devices such as optical computers, quantum simulators and inversionless lasers. Here we present a new approach to achieve strong coupling between electrically injected excitons and photonic bound states in the continuum of a dielectric metasurface monolithically patterned in the channel of a light-emitting transistor. Exciton-polaritons are generated by coupling electrically injected excitons in the gate-induced transport channel with a Bloch mode of the metasurface, and decay into photons emitted from the top surface of the transistor. Thanks to the high-finesse of the metasurface cavity, we achieve a large Rabi splitting of ~200 meV and more than 50-fold enhancement of the polaritonic emission over the intrinsic excitonic emission of the perovskite film. Moreover, we show that the directionality of polaritonic electroluminescence can be dynamically tuned by varying the source-drain bias which controls the radiative recombination zone of the excitons. We argue that this approach provides a new platform to study strong light-matter interaction in dispersion engineered photonic cavities under electrical injection, and paves the way to solution-processed electrically pumped polariton lasers.**

Exciton-polaritons are bosonic quasiparticles arising from the strong coupling between excitons and confined photons. The hybrid nature equips exciton-polaritons with extremely small effective mass, strong nonlinearities and fast relaxation,[1,2] properties that make them a unique platform for fundamental studies such as Bose-Einstein condensation and superfluidity, as well as for the realization of optoelectronic devices such as quantum simulators, neuromorphic computing and inversionless lasers.[3-12] While electrically injected polaritons could enable the realization of electrically pumped polariton lasers, polariton circuits and high-speed communication systems, so far polaritonic devices have mostly been driven by optical excitation.

The typical platform for studying electrically driven exciton polaritons is that of light emitting diodes (LEDs) integrated with distributed Bragg reflector (DBR) microcavities,[13-15] which has inherent limitations to further manipulation of the polaritonic emission. Dielectric metasurfaces, on the other hand, enable complete mode and dispersion engineering of the polaritons, but achieving electrically driven exciton-polaritons in metasurfaces is challenging due to the need of smooth contacts between the active materials and the charge injection layers to guarantee efficient charge carrier transport. By comparison, the lateral configuration of light-emitting transistor (LET) devices enables the fabrication of pixelated metasurfaces directly in the emission zone, with minimal degradation of device performance.[16] This provides the opportunity to design metasurfaces that support high Q-factor modes such as bound states in the continuum (BICs), which could ensure strong confinement of photonic modes in the active layer and lead to stronger light-matter interaction under electrical excitation.

In this work, we demonstrate electrically driven exciton-polaritons in a perovskite light-emitting transistor with metasurfaces directly patterned into the open surface between top electrodes. Metal halide perovskites are chosen as both, active medium and dielectric matrix due to their unique combination of high optical gain and luminescence quantum yield, good charge carrier transport, and high refractive index.[17-20] Importantly, excitons in perovskite can possess large binding energy, strong nonlinearities and large oscillator strength, which highly benefit the study of exciton-polaritons.[21-27] We observe the formation of high-quality factor polariton BICs (p-BICs) due to the strong coupling between the photonic BICs supported by the metasurface and the perovskite excitonic resonance, leading to a Rabi splitting energy of ~200 meV and more than 50-fold enhancement of the electroluminescence (EL) from the lower polariton band (LPB) over the intrinsic excitonic emission. Furthermore, we show that the directionality of the polaritonic emission can be dynamically tuned through group velocity selection under reversed electrical biases. Overall, these results build a foundation for a new class of light-emitting devices enabled by functional designer metasurfaces that operate in the strong-coupling regime, which could enable new fundamental discoveries and practical applications such as electrically driven polariton lasers and tunable perovskite polaritonic devices.

**Monolithic integration of metasurfaces into perovskite LETs**

The high refractive index of metal halide perovskites allows the monolithic integration of functional dielectric metasurfaces in light-emitting devices.[16,20,28-31] Here light-emitting transistors with lateral electrode configuration are employed for the accessibility of their active region to metasurface nanofabrication (Fig. 1). Methylammonium lead iodide (MAPbI$_3$) LETs were fabricated as detailed in the Methods section and elsewhere.[32-34] The transistor has a top-contact bottom-gate configuration, with gold (Au) and p-doped Si as the source-drain electrodes and gate electrode, respectively (Fig. 1a). To create the monolithic light emitting metadevice, or metatransistor,[32] the metasurface is directly patterned by focused ion beam (FIB) lithography (Fig. 1b). The metasurface, consisting of periodic nanobeams oriented parallel to the Au electrodes, serves as an optical cavity that supports photonic resonances. The highly confined cavity photons can strongly couple with excitons generated or injected in MAPbI$_3$ to form exciton-polaritons. The inset shows a microscope image of the operating metatransistor in which the bright square in the middle of the electrodes corresponds to the EL emission from the metasurface area. Figure 1c shows the low-temperature absorption and photoluminescence (PL) properties of MAPbI$_3$, where the absorption peak at 1.70 eV corresponds to the excitonic resonance of the perovskite, while the PL displays a dominant peak at ~1.66 eV and a minor peak at 1.60 eV, which can be attributed to the orthorhombic and tetragonal phases of MAPbI$_3$, respectively.[35]

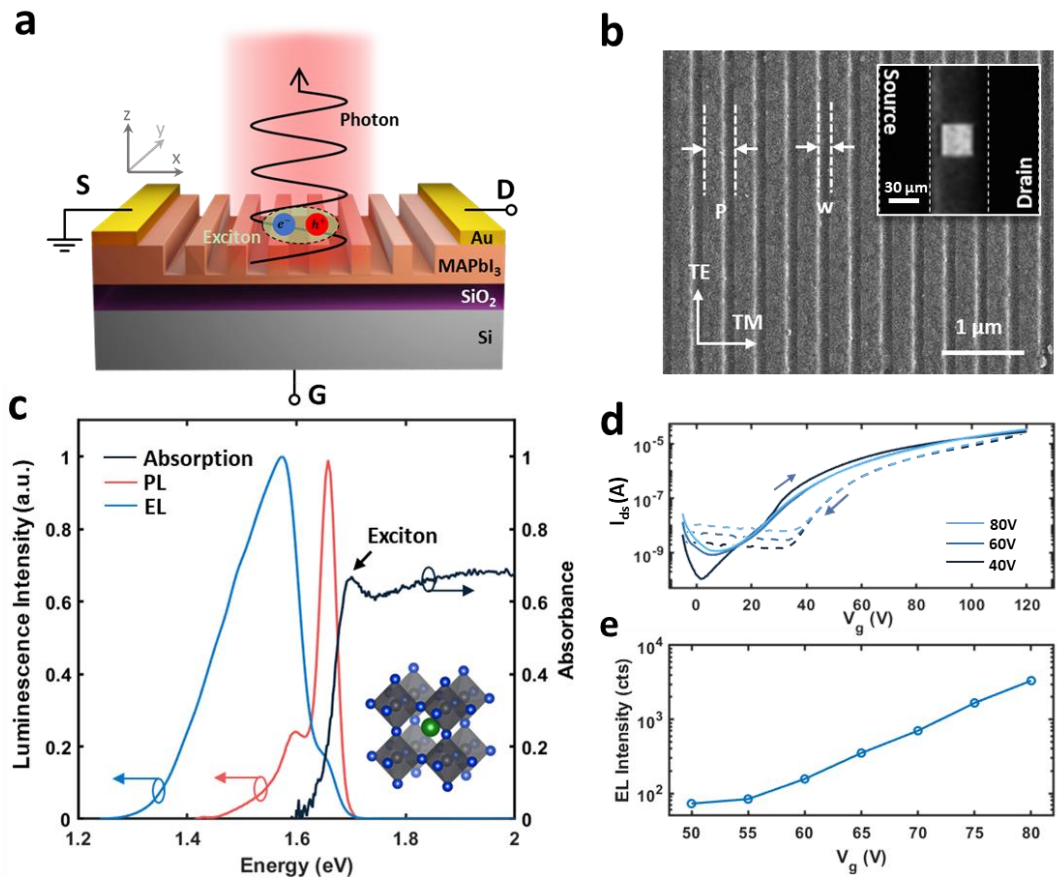

**Figure 1. Perovskite light-emitting metatransistors.** a) Schematic of the metatransistor device concept for polaritonic emission. A resonant dielectric metasurface is monolithically fabricated within the active region of the transistor between source (S) and drain (D) electrodes, leading to strong coupling of photonic modes and electrically injected excitons. The resulting polaritonic electroluminescence can be controlled by the gate (G) bias. b) Scanning electron microscope (SEM) image of the perovskite nanobeam array fabricated by FIB lithography within the active region between the top S-D electrodes. The perovskite nanobeams had period (P) ranging from 380 and 400 nm, height of 125 nm, and spacing (w) of 100 nm. The optical microscope image of the device operating at $V_g=\pm80$ V and $V_{ds}=80$ V in the inset shows the region of enhanced EL corresponding to the nanopatterned region of the square metasurface. c) PL and absorption spectra of the MAPbI$_3$ perovskite compared to the EL spectrum of the unpatterned LET at 78 K. The crystal structure of the MAPBI$_3$ perovskite is shown in the inset, where green, black and blue spheres represent MA$^+$, Pb$^{2+}$ and I$^-$ ions, respectively. d) Transfer characteristics of the MAPbI$_3$ metatransistor obtained in the forward and backward sweeping direction of gate bias with $V_{ds}=40$, 60 and 80 V. e) Peak EL intensity dependence on AC gate bias.

To increase brightness and uniformity of the electroluminescence, the device is operated under an alternating current (AC) gate voltage ($V_g$) of ±80 V and constant source-drain voltage ($V_{ds}$) of 80 V. Note that the relatively large bias voltages needed to operate the LETs are due to the large channel length and may be substantially reduced by decreasing the electrode gap or increasing the gate capacitance (e.g., adopting thinner $SiO_2$ layers or high-k dielectric materials).[36] The EL spectrum of the unstructured LETs peaks at 1.58 eV. The reduction of EL emission energy compared to the PL spectrum peaking at 1.66 eV may be due to energy transfer toward the smaller bandgap phase or emission from radiative trap states of the $MAPbI_3$ film upon electrical injection (Fig. 1c).[35,37] The gate bias dependence of the LET transport and EL characteristics are shown in Figures 1d and 1e. The perovskite metatransistor transfer characteristics show predominantly *n*-type charge transport behaviour with an on-off ratio larger than $10^3$ and electron mobilities up to ~0.16 $cm^2V^{-1}s^{-1}$ (Fig. 1d). The EL intensity is found to increase nearly exponentially when the gate voltage increases from 50 to 80 V (Fig. 1e). The strong dependence of the EL intensity on gate bias confirms that the device is operating in the enhancement regime.

**Electrically driven polariton BICs in perovskite light-emitting metatransistors**

To fulfil the criteria of strong coupling, the averaged decay rate of the excitonic ($\gamma$) and photonic resonances ($\kappa$) has to be smaller than the rate of the energy exchange ($g$) between them, i.e., $2g \geq (\kappa + \gamma)/2$ [21]. Since $\kappa$ is inversely proportional to the quality factor (Q) of the photonic cavity, a high Q cavity eases the strong coupling requirement. Here, we adopt a nanobeam metasurface that supports photonic BICs with extremely high radiative Q-factor to facilitate the formation of exciton-polaritons (Fig. 2). To understand the formation of BICs in the nanobeam structures, we simulated the photonic bands of the passive cavity (i.e., the perovskite replaced by a dielectric medium with a fixed refractive index n=2.4) with P=400 nm for TM and TE polarization, as shown in Figures 2a and 2b, respectively. Both TM and TE polarizations support symmetry-protected BICs at the Γ point located near 1.58 eV and 1.76 eV, respectively, indicated by the white arrow. The BICs are evidenced by the vanishing reflectance at the normal incidence, corresponding to the intrinsic near-zero linewidth and the ultrahigh radiative Q-factor of the passive metasurface (Fig. S1). The field distribution of the symmetry-protected BICs is shown in the insets, providing a visual clue that the BICs can hardly couple to the vertical plane wave due to symmetry mismatch[38].

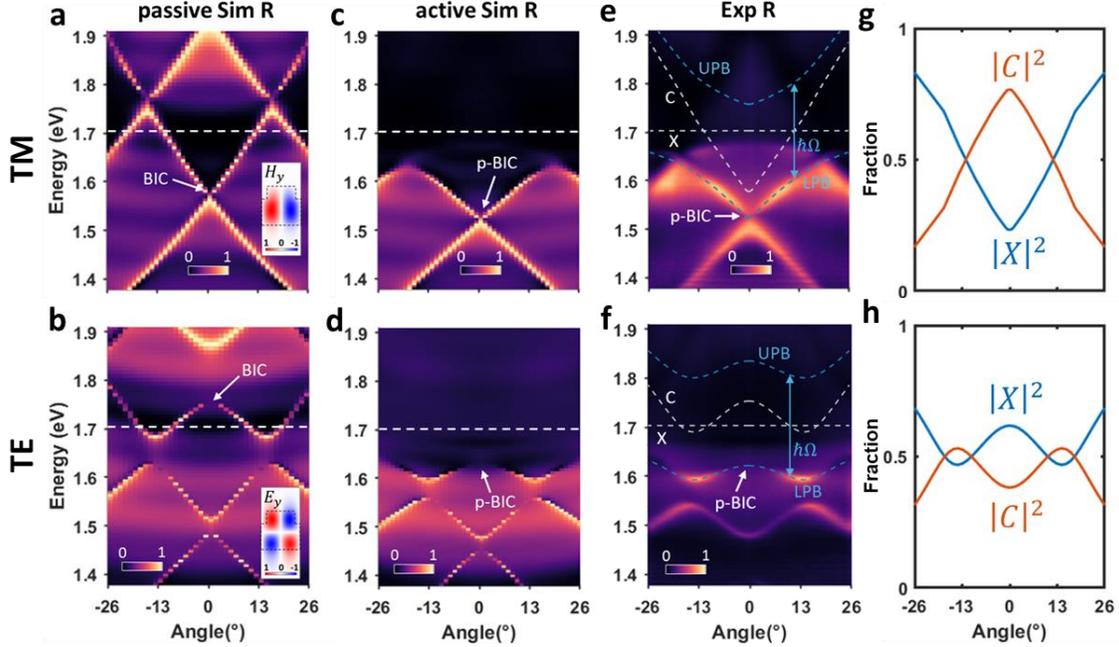

**Figure 2. Strong coupling of photonic BICs and perovskite excitonic resonances.** a, b) Calculated angle-resolved reflection spectra for TM (a) and TE (b) polarized illumination of the passive perovskite metasurface (i.e., assuming zero absorption and a fixed refractive index n=2.4). The white dashed lines indicate the position of the excitonic resonance of MAPbI$_3$ at ~1.70 eV. The insets show the field distribution in the *x-z* plane of the TM and TE polarized BICs. c, d) The simulated angle-resolved spectra of the active perovskite metasurface (P=400 nm) with TE and TM polarizations. e, f) Experimental angle-dependent reflection spectra of the active perovskite metasurface under TM and TE illumination, respectively. The white dashed lines indicate the band structure of the passive metasurface (C) and the excitonic resonance (X) of MAPbI$_3$. The blue dashed lines represent the theoretically fitted lower polariton (LPBs) and the upper polariton bands (UPBs). The polariton BICs are indicated by white arrows. g, h) Hopfield coefficients of the corresponding polaritons, showing exciton and photon fractions in the LPBs for TM and TE polarizations, respectively.

Angle-resolved reflection measurements are used to validate the coupling between the perovskite exciton resonances and the photonic BICs in the perovskite metasurface. The experiments were conducted at cryogenic temperature (i.e., T=78 K) using a home-built micro-spectrometer setup with back-focal plane capability[39] (details on the experimental setup are provided in Figure S2). The measured reflection spectra under TM and TE polarization excitation shown in Figures 3e and 3f are in excellent agreement with the simulated reflection band structures of the active metasurface (Fig. 3c and 3d) calculated by taking into account the dispersion of the refractive index (Fig. S3). Anticrossing behaviour between the photonic (C) and excitonic (X) bands can be observed for both TM and TE polarizations, a clear signature of strong coupling. A coupled oscillator model was used to calculate upper (UPB) and lower polariton branches (LPB)[2], showing good agreement with the reflection measurement results. Under cavity detuning of $\Delta_{TM}$=-126 meV ($\Delta_{TE}$=53 meV), the Rabi splitting energy is estimated to be 202 meV (206 meV) for the TM mode (TE mode). This is much larger than the averaged damping rates of the excitonic and photonic resonances (39 meV, as derived from the averaged linewidth), confirming the system is able to reach the strong coupling regime. Note that the upper polariton bands can hardly be distinguished in the experiment due to strong absorption above the exciton energy.[22,40] The Hopfield coefficients of the system, representing the fraction of the excitons ($|X|^2$) and the photons ($|C|^2$) in the LPBs for TM and TE polarizations, are calculated by the coupled oscillator model and shown in Figure 3g and 3h, respectively. Photons and excitons are equally mixed in the polariton branches at the angle where the passive photonic bands cross the excitonic resonance. In our case, at the Γ point the TM (TE) polarized polariton branch has an exciton fraction of 23% (62%), indicating the hybrid nature of the polariton BICs. The significant excitonic fractions of TE and TM polaritons in their LPBs are expected to facilitate polariton relaxation into lower energy states, which may eventually lead to condensation and inversionless lasing.[1,12,41,42]

Strong coupling of photonic BICs and excitonic resonances is also manifested by the polaritonic character of photoluminescence and electroluminescence properties of the perovskite metatransistor (Fig. 3). Angle-resolved photoluminescence spectra of the perovskite nanobeam array excited by an off-resonance laser show clear enhancement of the emissions from the LPBs, in good agreement with the theoretical predictions for both TM and TE polarizations (Fig. 3a and 3b). To demonstrate electrically driven exciton-polaritons in the perovskite metatransistor, the device was operated under AC gate voltage, and the angle-resolved EL spectra were recorded for TM and TE polarizations (Fig. 3c and 3d). Like PL, the EL spectra also show clear enhancement of polaritonic emission from the LPBs, confirming the generation of exciton-polaritons under electrical injection. It is worth noting that the outcoupling of the p-BICs at 1.53 eV for TM polarization (1.62 eV for TE polarization) to the radiative continuum is strictly forbidden due to the intrinsic symmetry mismatch inherited from the photonic BICs. The decrease of the intensity near the normal emission direction, therefore, is clear evidence of the p-BICs.

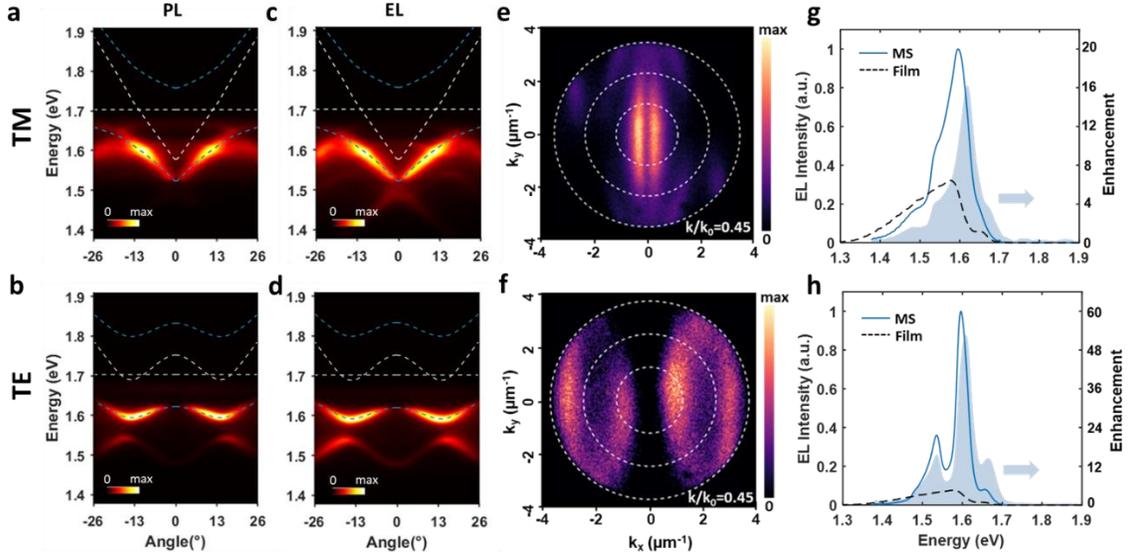

**Figure 3. Polaritonic emission from perovskite metatransistor.** a, b) Angle-resolved PL spectra of the perovskite metatransistor. c, d) Angle-resolved EL spectra of the perovskite metatransistor operating under $V_{ds} = 80V$, $V_g = \pm 80V$. e, f) k-space far-field profiles of the p-BICs, obtained with a 10-nm band pass filter centered at 810 nm (1.53 eV) for TM and 760 nm (1.63 eV) for TE polarization. The white dashed circles, from the inner to the outer, indicate the numerical aperture (NA) of 0.15, 0.3 and 0.45. g, h) EL intensity of the perovskite metasurface (blue line) and the perovskite film (black dashed line), along with the EL enhancement (blue shaded area) obtained as the ratio of the two.

The far field profile of the electrically driven p-BICs was analyzed in the xy-reciprocal space. Figures 3e and 3f show the back focal plane (BFP) image of the p-BICs around 1.53 eV (1.63 eV) for TM (TE) polarization. The EL at normal direction ($k_x=0$) is found to vanish even at large $k_y$ for both polarizations, matching well the simulated profile (Fig. S4). Interestingly, the polaritonic EL reaches a maximum 56-fold enhancement under TE polarization and 15-fold enhancement under TM polarization compared to the excitonic emission from the unstructured portions of the device channel, due to Purcell enhancement and scattering from the exciton reservoir into the LPB (Fig. 3g, h)[43]. Overall, the monolithically structured metatransistors yield higher emission directionality and higher density of states in the LPBs, thanks to the strong coupling. As discussed in the following, we believe that these characteristics provide new opportunities to realize highly efficient and tunable polaritonic devices and, potentially, electrically-driven polariton lasers.

**Electrically tunable directionality of polariton emission**

Unlike conventional LEDs where charge injection, and thus emission, is homogeneous across the active region of the device in the vertical multilayer stack, the recombination (emission) zone of LETs can be shifted towards the drain or the source electrode by adjusting the source-drain bias[32,44]. This characteristic feature of the LETs was exploited to prove reversible tunability of the polariton emission directionality in metatransistors (Fig. 4). Figure 4a shows the angle-resolved EL spectra of the perovskite metatransistor with detuning energy $\Delta_{TM}$=-73.1 meV (P=380 nm) with $V_{ds}$=80 V, which induces uniform EL emission throughout the metasurface area (inset of Fig. 4a). At these biasing conditions, two symmetric LPBs (LPB-1 and LPB-2) can be observed in the EL spectrum. Note that the enhanced polariton relaxation in LPB-1, characterized by the bottleneck where polaritons assemble, is located closer to the normal direction compared with the one presented in Figure 3c (i.e., $\Delta_{TM}$=-126.3 meV, P=400 nm) due to the larger exciton fraction and longer polariton lifetime at smaller detuning energy.[45] When reducing $V_{ds}$ to -40V, the emission zone shifts toward the source electrode, resulting in a gradient EL intensity distribution along the *x*-axis (inset of Fig. 4b). In this asymmetric electrical injection configuration, the majority of polaritons is driven to propagate along the positive *x* direction (with positive group velocity), as evidenced by the asymmetric EL bands of TM polarization in the experiment (Fig. 4b) and the extracted group velocity from the dispersion of LPB-1 and LPB-2 (Fig. 4c). The simulated angle-resolved emission using a dipole source on one side of the metasurface to mimic the gradient EL distribution shows good correspondence with the experimental results (Fig. S5). Specifically, for the polaritonic emission at 1.57 eV, exciton-polaritons are emitted at both 4° and -4° under symmetric injection ($V_{ds}$=80 V) while they can be electrically switched to 4° off the normal (with positive group velocity) by changing the source-drain bias to $V_{ds}$=-40 V (Fig. 4d). A similar phenomenon is also observed for TE-polarized polaritons (Fig. S6). This mechanism provides a simple way to electrically tune the directionality of the polaritonic emission, which may find applications in actively reconfigurable polaritonic devices. Furthermore, switching of the emission zone between the source and drain electrodes, which is possible in DC-driven LETs operating in the ambipolar regime,[16,36,44] would allow modulation of the polariton emission from positive to negative angles (i.e., with positive or negative group velocities), or at both angles with adjustable ratio, enabling pluridirectional beam steering and ranging.

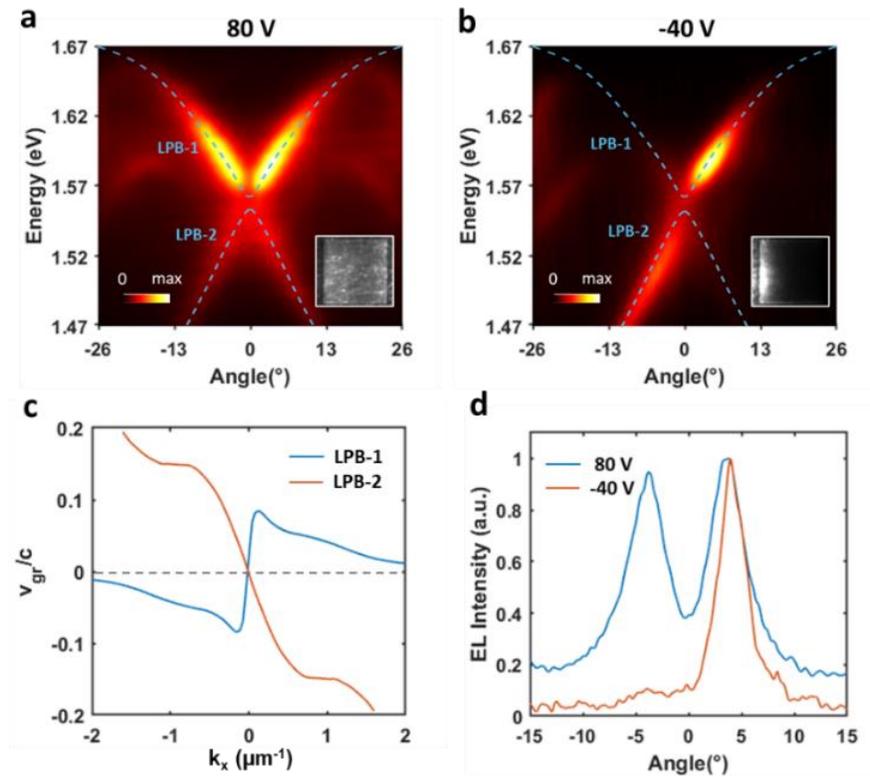

**Figure 4. Electrically tunable directional polaritonic emission.** TM polarized angle-resolved EL spectra of the perovskite metatransistor with P=380 nm, $V_g$ = ±80V and a) $V_{ds}$=80 V or b) $V_{ds}$=-40 V. Fitting of the LPB-1 and LPB-2 are indicated by the blue dashed lines. The insets show optical microscope images of the corresponding metasurface area

(30 μm x 30 μm). c) Polariton group velocity of the LPB-1 and LPB-2 extracted from the dispersion in a). d) Directional EL of LPB-1 at the energy of 1.57 eV obtained by applying 80 V (blue line) and -40V (orange line) source-drain voltage.

**Conclusions and outlook**

In conclusion, we demonstrated electrically driven exciton-polaritons in a MAPbI$_3$ light-emitting metatransistor with maximum Rabi splitting energy of 206 meV at 78 K. Low damping rates of the photonic BICs supported by the metasurface facilitate their strong coupling with the perovskite excitonic resonance to form polariton BICs under optical or electrical injection. This leads to more than 50-fold enhancement of EL from the LPB over the excitonic emission of the unpatterned MAPbI$_3$ film. Moreover, the directionality of the polaritonic emission can be reversibly tuned by controlling the position of the recombination zone within the transistor channel with different source-drain biases. Our proof-of-concept demonstration shows that the light-emitting metatransistor platform, which is compatible with various monolithic cavity designs and pixelated metasurface fabrication, enables complete polaritonic mode and dispersion engineering for the study of strong light-matter interactions under electrical excitation. To assess whether the threshold for polariton lasing is within reach, we estimated that the charge carrier density in the unoptimized MAPbI$_3$ LET device ($1.2\times10^{17}$ cm$^{-3}$) is only one order of magnitude lower than the threshold density obtained for photonic BIC lasing in MAPbI$_3$ metasurfaces ($1.6\times10^{18}$ cm$^{-3}$),[46] which may be even lower for polariton lasing (refer to "Charge carrier density estimation" section in Supplementary Information). Polariton condensation in perovskite light-emitting metatransistors could therefore be achieved by better cavity design, improved film and material quality, and higher injection currents (e.g. by optimized gate design and intense pulsed bias operation), opening new routes towards the highly sought, solution-processed electrically driven laser.[47,48]

**Methods**

*LET fabrication*

Perovskite LETs were fabricated in a bottom-gate and top-contact configuration on heavily *p*-doped Si substrates with a thermally grown SiO$_2$ (500 nm) layer as the gate insulator.

MAPbI$_3$ thin films were fabricated from a 1.3 M precursor solution of CH$_3$NH$_3$I (Dyesol) and PbI$_2$ (99.99%, TCI) (molar ratio 1:1) in anhydrous dimethylformamide (DMF, Sigma-Aldrich). The as-prepared solution was magnetically stirred for 2 hrs at 373 K in N$_2$ filled glovebox, then filtered by a polyvinylidene fluoride (PVDF) syringe filter (0.45 μm) before spin-coating. Prior to perovskite deposition, substrates were cleaned with ultrasonication for 15 min in acetone, isopropanol and deionized water. Subsequently, substrates were dried with the flow of nitrogen, followed by an oxygen plasma cleaning treatment. The perovskite precursor solution was spin-coated onto the quartz substrates with a speed of 4500 rpm for 30 s using the anti-solvent deposition method, with toluene drop-cast on the substrates 5 s after starting spinning. The resulting films were finally annealed at 373 K for 15 min. Afterwards, 100 nm thick Au electrodes were thermally evaporated in high vacuum (~10$^{-6}$ mbar). To avoid thermal decomposition of the perovskite films, samples were placed on a water-cooled substrate holder at 291 K during the electrode deposition. The resulting LET channel length (L) and width (W) were 80 μm and 1 mm, respectively.

*Metamaterial fabrication*

To complete the metatransistor devices, nanobeam arrays were patterned on the perovskite film by focused ion beam (Helios 600 NanoLab, FEI) milling to form the subwavelength grating metasurface. The lateral dimension of each fabricated metasurface was 30 μm×30 μm.

*Electrical and optical characterization*

Electrical and optical characterization was carried out using a temperature-controlled probe stage (HFS600E-PB4/PB2, Linkam) at the temperature of 78 K in the dark and under vacuum (~10$^{-3}$ mbar). DC characteristics were acquired with a 2-channel precision source/measure unit (B2902A, Agilent). Charge-carrier mobilities were extracted from the forward sweeping of transfer characteristics obtained at V$_{ds}$ = 40 V, using the conventional equations for metal-oxide semiconductor (MOS) transistors in the saturation regime: $\mu_{sat} = \frac{2L}{WC_i}\left(\frac{\partial\sqrt{I_{ds}}}{\partial V_g}\right)^2$.

Photoluminescence spectra were measured using a home-built setup based on an optical microscope (Eclipse LV100, Nikon) with LU plan fluor × 10 and × 50 objectives. The excitation source was a 405 nm picosecond laser diode (P-C-405B, Picoquant) operated at a 40 MHz repetition rate. The PL signal was detected by an AvaSpec ULS-RS-TEC, Avantes spectrometer.

Electroluminescence measurements of unpatterned devices were performed on the same optical setup as the PL measurement, under AC-driven mode by applying constant drain-source bias ($V_{ds}$=80 V) and square wave bias to the gate electrode (100 kHz modulation frequency), using an arbitrary waveform generator (3390, Keithley) coupled with a high-voltage amplifier (WMA-300, Falco Systems). Optical images and videos were taken and acquired by a cooled sCMOS scientific camera (PCO.edge 3.1m) coupled to the optical microscope.

*Angle-resolved measurements*

Angle-resolved reflection, PL, and EL measurements were performed with a home-built micro-spectrometer. The system consists of an inverted optical microscope (Nikon Ti-U), a spectrograph (Andor SR-303i with 150 lines/mm grating), and an electron-multiplying charged-coupled detector (EMCCD, Andor Newton 971). A lens system along the light path between the microscope and the spectrograph is used to project the back focal plane of the collection objective (Nikon ×50, with numerical aperture NA=0.45) onto the slit of the spectrograph. This configuration allows spectral measurement with angular information corresponding to the NA of the objective. A linear polarizer placed on the optical path of the lens system was used to select the collection polarization. A halogen lamp was used as an excitation source in reflection measurements, while in PL measurements, samples were excited by a 488 nm continuous wave solid state laser.

*Coupled oscillator model*

The coupling between photon and exciton resembles the coupling of two harmonic oscillators, so the system can be described by the coupled oscillator model. Assuming the photonic oscillator in the optical cavity possesses the resonance energy of $E_{cav}$ with the decay rate of $\gamma_{cav}$ while the excitonic resonance has the energy of $E_{exc}$ with a nonradiative decay rate of $\gamma_{exc}$, the Hamiltonian of this system can be written with a coupling strength of $2g_0$ (corresponding to the Rabi splitting energy $\hbar\Omega$) and the system can be described by the following eigenvalue equation:

$$\begin{pmatrix} E_{cav} + i\gamma_{cav} & 2g_0 \\ 2g_0 & E_{exc} + i\gamma_{exc} \end{pmatrix} \cdot \begin{pmatrix} C \\ X \end{pmatrix} = E \cdot \begin{pmatrix} C \\ X \end{pmatrix} \tag{1}$$

The C and X are the weighing factors of the photonic and excitonic oscillator, referring to the Hopfield coefficient. Let $\Delta E = E_{exc} - E_{cav}$, X and C are given by:

$$|X|^2 = \frac{1}{2}\left(1 + \frac{\Delta E}{\sqrt{\Delta E^2 + 4g_0^2}}\right) \tag{2}$$

$$|C|^2 = \frac{1}{2}\left(1 - \frac{\Delta E}{\sqrt{\Delta E^2 + 4g_0^2}}\right) \tag{3}$$

Solving the eigenvalue equation (1), gives the result of the eigenenergy:

$$E_{UP,LP} = \frac{1}{2}(E_{cav} + i\gamma_{cav} + E_{exc} + i\gamma_{exc})$$
$$\pm \sqrt{4g_0^2 + [E_{exc} + i\gamma_{exc} - E_{cav} + i\gamma_{exc} - i\gamma_{cav}]^2} \tag{4}$$

where, the $E_{UP}$ and $E_{LP}$ are the energy of the lower polariton state and upper polariton state, respectively. The formation of these two new eigen modes indicates a clear anticrossing behavior. Assuming a non-dispersive $E_{exc}$ with regard to the in-plane wave vector $k_\parallel$, the dispersion of the polariton bands can be determined by the dispersion of the photonic bands $E_{cav}(k_\parallel)$. The coupling strength between the two oscillators must be larger than the averaged decay rates in order to observe the splitting of the lower polariton state and the upper polariton state:

$$2g_0 > \frac{(\gamma_{cav} + \gamma_{exc})}{2} \tag{5}$$

*Numerical simulations*

The angle-resolved reflection spectra were computed in Comsol Multiphysics 5.4. Floquet periodic boundary conditions were used in the transverse direction, and perfectly matched layers were used in the propagation direction. The system was excited by a periodic port above the array. The far-field distribution of the emission in the momentum space was calculated by the reciprocity method, i.e., integrating the electric field intensity ($\sim|E|^2$) inside the perovskite grating metasurface for different angles of incidence. The optical bands and Q-factors were calculated by an eigenfrequency solver using the same model.

**Acknowledgements**

We thank Dr. Ramón Paniagua-Domínguez and Dr. Arseniy I. Kuznetsov for valuable discussions and advice on this work. Research was supported by the Agency for Science, Technology and Research A*STAR-AME programmatic grant on Nanoantenna Spatial Light Modulators for Next-Gen Display Technologies (Grant no. A18A7b0058) and the Singapore Ministry of Education MOE Tier 3 (Grant no. MOE2016-T3-1-006). S.T.H acknowledges support from AME Young Individual Research Grant (YIRG Grant no. A2084c0177).

**Author contributions**

C.S., S.T.H. and Y.W. conceived the idea. Y.W. and M.K. fabricated the perovskite LETs. G.A. carried out FIB lithography for metasurface fabrication. Y.W. characterized PL, EL and absorption of the devices with M.K., conducted numerical simulations of the metasurface response and theoretical analysis with J.T.. S.T.H. carried out angle-resolved reflection, PL and EL measurements with assistance from Y.W. and M.K.. Y.W, J.T., S.T.H. and C.S. performed data analysis and wrote the paper with inputs from all authors.

**Referencees**

# Supplementary Materials for

**Electrically driven exciton-polaritons
in metal halide perovskite metatransistors**

*Yutao Wang,[1,2†] Jingyi Tian, [1,3†] Maciej Klein, [1,3] Giorgio Adamo, [1,3]*

*Son Tung Ha,[4*] and Cesare Soci[1,3*]*

[1] Centre for Disruptive Photonic Technologies, TPI, Nanyang Technological University, 21 Nanyang Link, Singapore 637371

[2] Interdisciplinary Graduate School, Energy Research Institute @NTU (ERI@N), Nanyang Technological University, 50 Nanyang Drive, Singapore 637553

[3] Division of Physics and Applied Physics, School of Physical and Mathematical Sciences, Nanyang Technological University, 21 Nanyang Link, Singapore 637371

[4] Institute of Materials Research and Engineering, Agency for Science Technology and Research (A*STAR), 2 Fusionopolis Way, Singapore 138634

[†]These authors contributed equally to this work

*Correspondence: Ha_Son_Tung@imre.a-star.edu.sg; csoci@ntu.edu.sg*

**Supplementary Text**

**Charge carrier density estimation**

To evaluate the ability of electrically pumped lasing in perovskite meta-transistors, we first estimate the charge carrier densities required for optically pumped photonic lasing in the MAPbI$_3$ perovskite metasurface in another work of our group with the same fabrication procedure[1]. Considering the carrier lifetime in perovskite ranges is roughly 100 ns,[2] as the optical pump pulse duration (100 fs) used in the optical experiment is much smaller than the photogenerated carrier lifetime ($\tau_{pulse} \ll \tau_{carrier}$), carrier recombination can be reasonably neglected within a single pulse duration. Thus, assuming 100% carrier generation efficiency, the carrier density induced by the optical pump can be calculated by the following equation:[2]

$$n = (1-R)(1-e^{-\alpha L})(I_p/L \cdot \frac{hc}{\lambda})$$

where $R$ and $\alpha$ are the reflectance and absorption coefficients, $L$ the optical penetration depth, and $I_p$ is the laser pulse fluence. The corresponding carrier density at the threshold fluence of 8 uJ/cm$^2$ of the optically pumped BIC laser is $n \sim 1.6 \times 10^{18}$ cm$^{-3}$.

For electrically driven LETs under continuous operation, the carrier density is given by:

$$n = \frac{n\tau_c}{qd}$$

where $j$ is the current density, $\tau_c$ is the carrier lifetime, and $d$ is the transistor channel length. The current density achieved experimentally in our perovskite LETs has reached the level of ~1.5 kA/cm$^2$ according to the electrical characteristics. Assuming the carrier lifetime to be $\tau_c \sim 100$ ns, this corresponds to a carrier density of $n \sim 1.2 \times 10^{17}$ cm$^{-3}$. Since the threshold to achieve polariton condensation in the strong coupling regime is generally much smaller than the photonic lasing threshold,[3,4] it is promising to achieve electrically driven lasers based on perovskite meta-transistors by better cavity designs, improved quality of the gain materials, optimized contact geometry and higher injection currents.

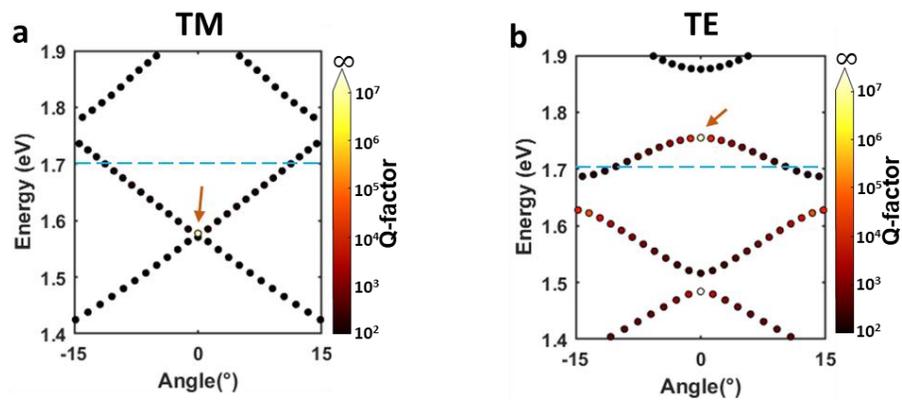

Figure S1. Calculated band structures and the Q-factors of the passive perovskite metasurface of a) TE, and b) TM polarization using eigenmode simulation. The orange arrows indicate the symmetry protected photonic BICs located at the Γ point with ultrahigh Q-factors. The excitonic resonance is indicated by the blue dashed line.

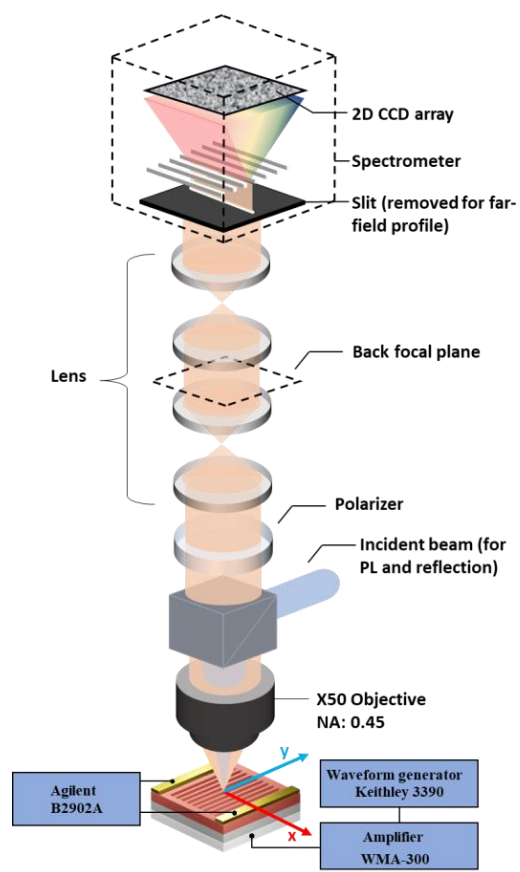

Figure S2. Schematics of the back focal plane imaging systems for angle-resolved reflection, PL and EL measurement. The slit is removed and a band pass filter is added after the polarizer for far field profile measurement.
21

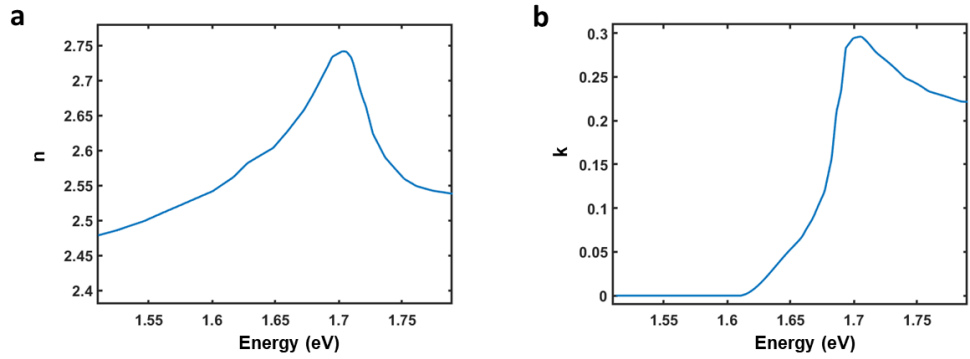

Figure S3. The a) real refractive index and b) extinction coefficient of the spin coated MAPbI$_3$ film at 78K. The refractive index (n=2.4) for passive metasurface simulation is estimated at low photon energy where the exciton oscillator has minor effect on the refractive index.

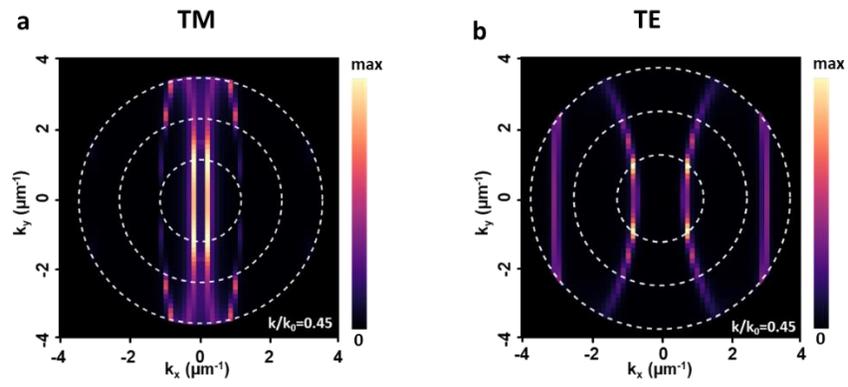

Figure S4. The simulated far field profile of the polariton BIC for a) TM and b) TE mode. The white dashed circles represent the NA of 0.15, 0.3 and 0.45, from inner to outer side.

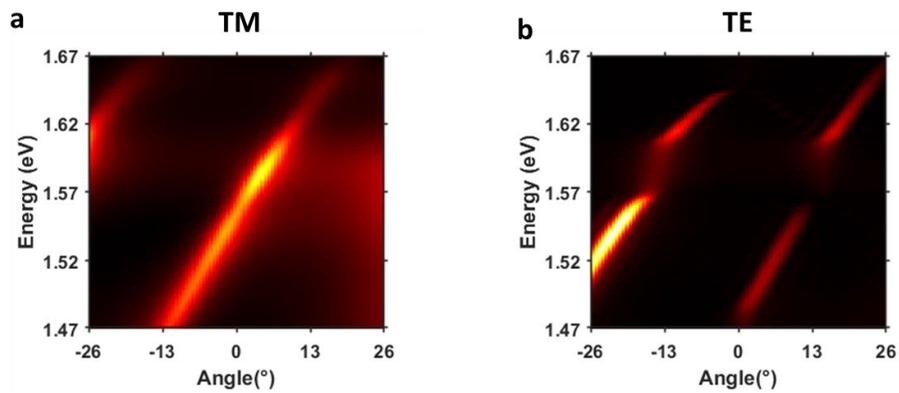

Figure S5. The simulated band structure of the a) TM and b) TE polarization with an asymmetric pumping situation. Similar to the experimental results, only mode with positive group velocity is emitting in the polariton bands. The asymmetric band structure is simulated by putting a x (TM) or y (TE) oriented electric dipole on one side of the nanogratings.

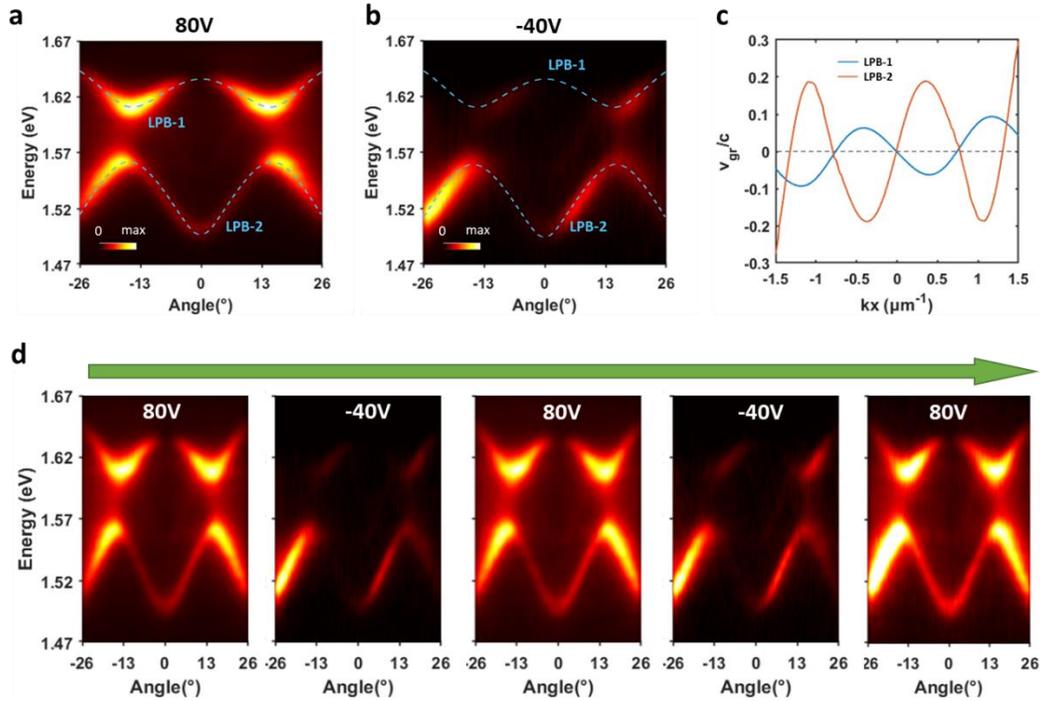

Figure S6. The angle resolved EL spectra of the TE polarization under a) symmetric ($V_{ds}$ = 80V) and b) asymmetric ($V_{ds}$ = -40V) pumping scheme. Blue dashed lines are the calculated polariton bands via coupled oscillator model. Two polariton bands (LBP-1 and LPB-2) can be observed. c) Group velocity of the LPB-1 and LPB-2. Similar to the case in Figure 4, only modes with positive group velocities are emitting in the asymmetric pumping scheme. d) Angle resolved EL spectra showing reversible switching between symmetric and asymmetric emission band structures.

25